\documentclass[preprint2]{aastex}

\usepackage{graphicx}
\usepackage{subfigure}
\usepackage{amsmath}
\usepackage{amsfonts}
\usepackage{amssymb}
\usepackage[T1]{fontenc}
\usepackage{calligra}

\def\be{\begin{equation}}
\def\ee{\end{equation}}

\def\ep{\epsilon}
\def\beq{\begin{eqnarray}}
\def\eeq{\end{eqnarray}}

\begin{document}

\title{Stars and Black Holes in Massive Gravity} 

\author{Andrei Gruzinov, Mehrdad Mirbabayi}

\affil{CCPP, Physics Department, New York University, 4 Washington Place, New York, NY 10003}

\begin{abstract}

Generically, massive gravity gives a non-unique gravitational field around a star. For a special family of massive gravity theories, we show that the stellar gravitational field is unique and observationally acceptable, that is close to Einsteinian. The black hole solutions in this family of theories are also studied and shown to be peculiar. Black holes have a near-horizon throat and the curvature diverging at the horizon.

We show that there exists a sub-family of these massive gravity theories with  non-singular at horizon black holes. 

~~

~~

\end{abstract}

\section{Introduction} 

The history of massive gravity, as it relates to the present work, can be summarized as follows. 

Fierz and Pauli (1939) introduce linearized massive gravity.

Van Dam and Veltman (1970) and Zakharov (1970) show that the Fierz-Pauli linearized massive gravity is strongly ruled out by the solar system (and many other) observations -- the so-called vDVZ discontinuity.

Vainshtein (1972) argues that, due to non-linear effects, massive gravity with small graviton mass is in fact observationally acceptable -- it gives close to Einsteinian gravitational field near massive enough bodies. 

The Vainshtein (1972) argument is not a real proof. Vainshtein just points out that a near-Einsteinian solution of a spherically symmetric static problem exists in a range of radii. But the existence of an acceptable solution in a finite range of radii does not, generically, guarantee that the same solution has the right asymptotic behavior at large and small radii. And indeed, Babichev et. al. (2010) show that the spherically symmetric solution in massive gravity is not even unique. 

A non-unique gravitational field of a massive body is strange and might be taken as a sign that  Fierz-Paulian massive gravities (all possible nonlinear completions of the original Fierz-Pauli action, call it FP) are sick. In fact ambiguous stellar gravity is not even the main problem with FP. FP is known to have ghosts (Boulware, Deser 1972) and unacceptable cosmologies (Gabadadze, Gruzinov 2005). 

It is not clear how general are the no-go results of Boulware and Deser (1972) and of Gabadadze and Gruzinov (2005). Maybe one can find some  very special FP theory which is free of ghosts and has an acceptable cosmology. An important recent finding (De Rham et al 2010) is that there exists just a two-parameter sub-family of FP theories which might be ghost free -- call it FP2 (Appendix A). All other FP theories have ghosts. 

In this paper we show that 
\begin{itemize}
\item FP2 gives a unique spherically symmetrical gravitational field for a given star or black hole.
\item Some FP2 theories give observationally acceptable gravitational field for stars.
\item Black holes in generic FP2 are peculiar -- they have a throat and a singular horizon.
\item There is a one-parametric sub-family of FP2 -- call it FP1 -- which gives black holes with non-singular at horizon (and close to Schwarzschild for small graviton mass) geometry, although there still exists a scalar quantity which is singular at horizon.
\end{itemize}

While the status of both FP2 and FP1 is almost as questionable as that of FP (being FP2 is a necessary condition for ghost-freedom, but it is not known if it is also sufficient), we present our results hoping that they might be qualitatively useful for any future theory of massive gravity. If the FP2 family does contain a good theory of massive gravity, then we tentatively predict that this good theory is from the FP1 sub-family.

\section{Uniqueness of the static spherical solutions in FP2}

Representing the contribution of the FP action terms by an additional stress-energy tensor in the Einstein equation, we write the spherically symmetric solution in the standard form 
\be \label{m1}
ds^2=e^\nu dt^2-e^\lambda dr^2-r^2d\Omega ^2.
\ee
One has to introduce an additional unknown (Vainshtein's $\mu (r)$) to describe the FP stress-energy. In FP2, however, the resulting system of equations is of lower order than in the generic FP. As shown in Appendix B, the vacuum equations can be written as
\be \label{e21}
\nu '=f_1(r, \nu ,\lambda ),
\ee
\be \label{e22}
\lambda '=f_2(r, \nu ,\lambda ),
\ee
where the functions $f_1$, $f_2$ are given (implicitly) in Appendix B. 

Equations (\ref{e21},\ref{e22}) require two integration constants, in contrast to GR where the $\lambda$-equation decouples and only one integration constant is needed. However, the condition of asymptotic flatness ($\nu \to 0$, $\lambda \to 0$ for $r \to \infty$) fixes one of the integration constants. Then, just like in Einstein theory, one gets a one-parameter family of vacuum solutions. This remaining parameter (mass in the Einstein theory) characterizes the central object.

\section{Stars in FP2}

At $r\gg m^{-1}$, we start with the linearized asymptotically flat solution (given in Appendix B) and numerically integrate the system (\ref{e21}, \ref{e22}) inwards. We modify the equations (\ref{e21}, \ref{e22}) appropriately after we cross the surface of the star in order to include the stress-energy of matter (Appendix C). For a given stellar radius and mass (as determined by the large-$r$ asymptotic), the correct value of the stellar energy density $\epsilon$ is obtained by numerical shooting. We use a simple matter equation of state $\epsilon ={\rm const}$.

If the gravity parameters are chosen appropriately \footnote{ As seen from the Bianchi identity (\ref{ebn}), to recover the linearized Schwarzschild solution, one needs to choose the gravity parameter $c_3>0$, in agreement with the decoupling limit analysis of Chkareuli and Pirtskhalava (2011). A careful study of the parameter space $(c_2, c_3)$ was not performed. In what follows, $c_3>0$ and $|c_2|$ is not too big.}, and so long as the graviton mass is small enough, the star solution agrees with the Einstein theory at small radii, and it agrees with the linearized massive gravity result at large radii (Fig. 1).

\begin{figure}[h!]
\plotone{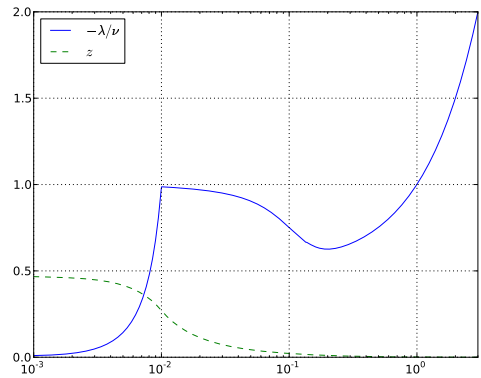}
\caption{\footnotesize Semi-logarithmic plot of $-\lambda/\nu$ and redshift $z=e^{-\nu/2}-1$ as a function of $r$. $-\lambda/\nu$ linearly grows with $r$ at large values of $r$ in agreement with the linearized massive gravity solution \eqref{linear}. Einstein theory is recovered at small radii since $-\lambda/\nu$ approaches the Schwarzchild metric value (i.e. $\lambda/\nu=-1$) outside the star, and $\lambda\to 0$ at the center. The surface of the star is located at $r_s=0.01$ and its mass is determined implicitly by $\nu=-0.005e^{-mr}/r$ at large $r$. The gravity parameters are $m=1$, $c_2=0$, $c_3=2$. }
\end{figure}


\section{Black Holes in FP2} 

For black holes, equations (\ref{e21}, \ref{e22}) cannot be integrated as written, because the solution features a coordinate singularity $w\equiv e^{-\lambda /2}=0$ at a finite redshift $z\equiv e^{-\nu/2}-1$. 

This means that the black hole develops a throat -- as one moves inward to larger $z$, the circumference of the sphere $z={\rm const}$  first decreases but then increases again. The solution then cannot be calculated with the coordinatization (\ref{m1}). 

We changed the coordinatization to
\be
ds^2=e^\nu dt^2-d\rho^2-r^2d\Omega ^2,
\ee
where $\rho$ is the new independent variable, while $\nu= \nu (\rho )$ and $r=r(\rho )$ are the new unknowns. We re-write the system (\ref{e21}, \ref{e22}) as (prime now denotes the $\rho$-derivative)
\be 
r'=w
\ee
\be 
w '=\tilde{f}_1(r, w, z),
\ee
\be 
z '=\tilde{f}_2(r, w, z).
\ee

These equations can be integrated all the way to the horizon $z=\infty$. The results are shown in Fig.2. As seen, one does recover the Schwarzschild metric at small radii, except in the near-horizon region. Near the horizon, the black hole develops a throat, and the curvature diverges at the horizon.

These features can be understood, in the small graviton mass limit, through an approximate near-horizon solution (Appendix D). Measuring $\rho$ from the point of minimum circumference $2\pi r_{\min}$, we get the (circumference-defined) radius $r$ and the Ricci curvature invariant $W\equiv \sqrt{R_{\mu \nu}R^{\mu \nu}}$: 
\beq
&r \simeq r_{\min}+\rho^2/(4r_{\min})\,,&\\
&W \simeq C_1 m^2 z = C_2 m^2/(\rho/r_{\min} +C_3m^2r_{\min}^2)\,&\\
&\rho/r_{\min}>-C_3 m^2r_{\min}^2
\eeq
Here $C_i$ are positive dimensionless numbers which depend on the dimensionless parameters of the theory $c_2$ and $c_3$. As $\rho$ decreases from positive to negative values, the circumference reaches a minimum value outside the horizon. Then one approaches the horizon, $z=\infty$, and $W$ diverges.

As shown in Appendix D, All $C_i$ actually vanish for $c_3=c_2^2$. This defines FP1 -- a one-parameter sub-family of FP2.

\begin{figure}[t]
\plotone{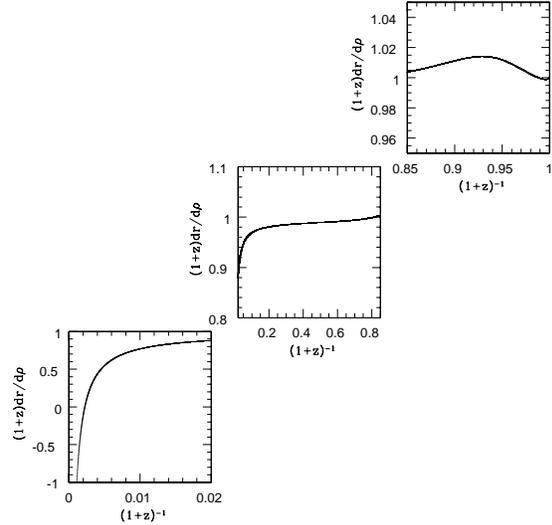}
\caption{\footnotesize $(1+z)dr/d\rho$ as a function of $(1+z)^{-1}$. Here $z$ is the redshift, $r$ is the radius as measured by the circumference of a constant-redshift sphere, $\rho$ is the proper radial distance. For the Schwarzschild metric, $(1+z)dr/d\rho =1$, which is indeed seen at intermediate redshifts. The negative values of $(1+z)dr/d\rho$ near the horizon ($z=\infty$) correspond to a throat. The gravity parameters are $m=1$, $c_2=0$, $c_3=2$. The mass of the black hole is given implicitly by $\nu=-0.06e^{-mr}/r$ at large $r$.}
\end{figure}

\section{FP1}

We do not understand what makes the FP1 (FP2 with $c_3=c_2^2$, see Apendices A, B for the definitions) special in any intuitive way. But we do understand which special property of the FP1 family is responsible for the non-singular black hole horizon.

The FP1 family is distinguished by the existence of a unitary-gauge orthogonally-coordinatized de-Sitter (and Schwarzschild-de Sitter) solutions, Nieuwenhuizen (2011). Namely, FP1  has solutions:
\be
ds^2=Adt^2-A^{-1}dr^2-Br^2(d\theta ^2+\sin ^2\theta d\phi ^2),
\ee
\be
A=B^{-1}(1-{r_g\over r}-{B m^2\over 3c_2}r^2),~~B={c_2^2\over (1+c_2)^2},
\ee
\be 
\phi ^\mu = x^\mu = (t, r\sin \theta \cos \phi , r\sin \theta \sin \phi ,r\cos \theta ).
\ee
Here the Schwarzschild radius $r_g$ is arbitrary, $m$ is the graviton mass, and $c_2$ is the dimensionless parameter of the FP1 family.

The existence of the exact Schwarzschild-de Sitter solution is of no direct interest for the present study -- we are only interested in the asymptotically flat solutions. But, as shown in Fig.3, in FP1 the asymptotically flat solutions switch to the exact Schwarzschild-de Sitter solution at redshift \footnote{ The solution shown in Fig.3 was obtained as the $c2=\sqrt{c_3}-0$ limit of the generic FP2 solution. For exactly FP1, the radius corresponding to $z=1/ c_2$ becomes a bifurcation point, i.e. at smaller radii ($z>1/ c_2$) one can either stay on the Schwarzschild-de Sitter branch $\lambda_2=-1/c_2$, or proceed to even lower values of $\lambda _2$. The latter behavior can be enforced by canceling the common factor $(1+c_2\lambda _2)$ from both sides of the Bianchi identity (\ref{ebn}). This branch has infinite curvature at horizon.}
\be 
z={1\over c_2}.
\ee

\begin{figure}[t]
\epsscale{1.1}
\plotone{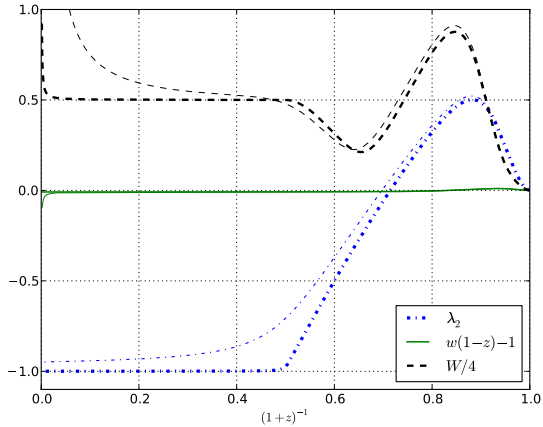}
\caption{\footnotesize  $\lambda _2$, $w(1+z)-1$, indicating closeness to Schwrazschild, and  $\sqrt{R_{\mu \nu}R^{\mu \nu}}/4$ are shown  vs $(1+z)^{-1}$. Thin: $c_2=0.9$, thick: $c_2=0.999$. Other gravity parameters:  $m=1$, $c_3=1$. The mass of the black hole is given implicitly by $\nu =-0.04e^{-mr}/r$ at large $r$.}
\end{figure}

\section{Conclusion}
There is a one-parameter family of Fierz-Paulian gravities, FP1,  which might have ``good'' black holes. As shown in Fig.3, these black holes are still strange. The spacetime switches to exact Schwarzschild-de Sitter at a finite redshift. As far as we can see, however, these black holes are the only non-singular black holes from the FP2 family of De Rham et al (2010). Given that FP2 is the only candidate ghost-free Fierz-Paulian gravity, these strange black holes might deserve further studies.

\section*{Acknowledgments}

We thank L. Berezhiani, G. Chkareuli, S. Dubovsky, G. Gabadadze, S. Mukhanov, and D. Pirtskhalava for useful discussions.

\renewcommand{\theequation}{A\arabic{equation}}
\setcounter{equation}{0} 

\section*{A. Massive gravity}

Massive gravity can be represented as Einstein gravity interacting with 4 non-canonical scalar fields (Arkani-Hamed et al 2003, Dubovsky 2004, Chamseddine, Mukhanov 2010). The action is the Einstein-Hilbert action plus the (non-linear version of) Fierz-Pauli action: 
\beq
&S=S_{EH}+S_{FP}, &\\
&S_{EH}=-{1\over 2}\int d^4x\sqrt{-g}R,& \\
&S_{FP}=m^2\int d^4x\sqrt{-g}U.&
\eeq
Here $U$ is an arbitrary symmetric function of the eigenvalues of the matrix
\be
\label{H}
H^A_B=\eta _{BC}g^{\mu \nu}\partial _\mu\phi ^A\partial _\nu\phi ^C,
\ee
where $g_{\mu \nu}$ is the metric tensor, $\eta _{AB} \equiv {\rm diag} (1,-1,$ $-1,-1)$ is the Minkowski matrix, and $\phi ^A$, $A=0,1,2,3$,  are the 4 scalar fields.
 
Since the action is generally covariant, one is free to choose an arbitrary coordinatization. In particular, for $x^\mu = \phi ^\mu$, and for the appropriate function $U$, one recovers the Fierz and Pauli (1939) mass term.

De Rham et al (2010) show that a two parameter family of massive gravities has some good features (ghost-free in the so-called decoupling limit). The family is described by the potential (Nieuwenhuizen 2011, Hassan, Rosen 2011, Koyama et al 2011)
\be \label{dR}
U=\sum \lambda _A\lambda_B+\tilde{c}_2\sum \lambda _A\lambda _B\lambda _C+\tilde{c}_3\lambda _0\lambda _1\lambda _2\lambda _3,
\ee
where the sums are over all all-distinct pairs and triples of indices, and $\lambda _A$ are the four eigenvalues of the matrix
\be
\delta ^A_B - \sqrt{ H^A_B }.
\ee

We call this two-parameter family of theories FP2. 

\renewcommand{\theequation}{B\arabic{equation}}
\setcounter{equation}{0} 

\section*{B. Static spherically symmetric field}

Starting from the unitary gauge $x^\mu=\phi^\mu$, the asymptotically flat metric can be written as (Appendix E)
\be 
ds^2=e^\nu dt^2-e^{\tilde{\lambda }}dR^2-R^2e^\mu d\Omega ^2.
\ee
We then change the radial coordinate to transform the metric into the usual form
\be 
ds^2=e^\nu dt^2-e^\lambda dr^2-r^2d\Omega ^2.
\ee
The scalar fields in the new coordinates become
\be
\label{phi}
\phi^0=t\,,\qquad \phi^i=re^{-\mu/2}n^i\,,
\ee
where $n^i$ is the unit radial vector. 

As independent equations we use two Einstein equations and the stress-energy conservation
\be \label{e0}
G^0_0=r^{-2}(1-e^{-\lambda})+r^{-1}e^{-\lambda}\lambda '=T^0_0, 
\ee
\be \label{e1}
G^1_1=r^{-2}(1-e^{-\lambda})-r^{-1}e^{-\lambda}\nu '=T^1_1, 
\ee
\be \label{eb}
{T^1_1} '={1\over 2}\nu '(T^0_0-T^1_1)+{2\over r}(T^2_2-T^1_1),
\ee
where prime denotes the $r$-derivative.

The stress-energy tensor of FP2 is derived from eq.\eqref{dR}: 
\be \label{t0}
T^0_0=-m^2(\lambda _1+2\lambda_2+c_2(2\lambda _1\lambda _2+\lambda_2^2)+c_3\lambda_1\lambda_2^2),
\ee
\be \label{t1}
T^1_1=-m^2(\lambda _0+2\lambda_2+c_2(2\lambda _0\lambda _2+\lambda_2^2)+c_3\lambda_0\lambda_2^2),
\ee
\beq \label{t2}
 T^2_2=-m^2(\lambda _0+\lambda_1+\lambda_2  +\\ 
 c_2(\lambda _0\lambda _1+\lambda _0\lambda _2+\lambda _1\lambda _2)+c_3\lambda_0\lambda_1\lambda_2), \nonumber
\eeq
with 
\be
c_2=1+\tilde c_2,~ c_3=\tilde c_2 +\tilde c_3,
\ee
and
\be 
\label{l0}
\lambda _0=1-e^{-\nu/2},
\ee
\be
\label{l1}
\lambda _1=1-e^{-\tilde{\lambda }/2}=1-e^{-\lambda/2}(1-\lambda_2-r\lambda_2'),
\ee
\be
\label{l2}
\lambda _2=1-e^{-\mu/2}.
\ee

We should remark that  $e^{-\mu/2}$ in (\ref{phi},\ref{l2}) and $e^{-\tilde{\lambda}/2}$ in \eqref{l1} are defined as square roots of $e^{-\mu}$ and $e^{-\tilde{\lambda}}$ and can therefore be negative, because the square root function must be allowed to switch to negative branch at zero as continuity dictates. We will ultimately work in terms of $\lambda_2$ instead of $\mu$ and $w\equiv e^{-\lambda/2}$, which automatically selects the right sign of the square root.

Using expressions (\ref{t0},\ref{t1},\ref{t2}) and expression (\ref{l0},\ref{l1}) in eqs.(\ref{e0},\ref{e1},\ref{eb}) gives the following system of three equations for three unknowns $\nu$, $\lambda$, and $\lambda _2$:
\beq \label{e0n}
1-e^{-\lambda}+re^{-\lambda}\lambda '=-m^2r^2( 2\lambda _2 
+ c_2\lambda_2^2+  \nonumber \\ 
(1+2c_2\lambda _2+c_3\lambda_2^2)(1-e^{-\lambda /2}(1-\lambda_2-r\lambda_2'))) \nonumber \\
\eeq
\beq \label{e1n}
1-e^{-\lambda}-re^{-\lambda}\nu '=-m^2r^2( 2\lambda _2+c_2\lambda_2^2+ \nonumber \\ 
(1+2c_2\lambda _2+c_3\lambda_2^2)(1-e^{-\nu /2}) )~~~
\eeq
\beq \label{ebn}
r\nu '(1+2c_2\lambda _2+c_3\lambda_2^2)=4(e^{\lambda /2}-1)( 1+c_2\lambda_2+\nonumber \\ 
 (c_2+c_3\lambda _2)(1-e^{-\nu /2}) ).~~~
\eeq

After linearizing in $\nu$, $\lambda$, $\mu$, equations (\ref{e0n}, \ref{e1n}, \ref{ebn}) can be solved exactly:
\beq 
\label{linear}
\nu =-ce^{-mr}{1\over r}, ~~\lambda ={c\over 2}e^{-mr}({1\over r}+m),\\ \nonumber 
\mu = {c\over 2m^2}e^{-mr}({1\over r^3}+{m\over r^2}+{m^2\over r})\,,
\eeq
manifesting the vDVZ discontinuity. 

A Vainshtein type argument can  be used at this point. Neglect the graviton mass in eqs. (\ref{e0n},\ref{e1n}) to recover the Einstein theory. Then substitute the linearized Schwarzchild solution $\nu=-\lambda=-r_g/r$ into the Bianchi identity \eqref{ebn}, also linearized in $\nu$ and $\lambda$, but exact in  $\lambda_2$, and solve for $\lambda_2$. One finds
\beq
c_3\lambda_2^2=1\,.
\eeq
which implies $c_3>0$.

Without linearizing, equations (\ref{e0n}, \ref{e1n}, \ref{ebn}) cannot be solved as written. Although we do have three equations for the three unknowns, it is seen that the only derivatives of the unknowns in these equations are $\lambda '$ and $\lambda _2 '$ in eq.(\ref{e0n}), $\nu '$ in eq.(\ref{e1n}), and $\nu '$  in eq.(\ref{ebn}). 

Equating the two expressions for $\nu '$, we derive an algebraic relation between $r$, $\nu$, $\lambda$, $\lambda _2$. Then one can select any two of the three unknowns $\nu$, $\lambda$, $\lambda _2$ and derive a system of two first-order differential equations for the two selected unknowns.

In practice\footnote{ The results of the following calculation are to be used by a computer. Therefore, we organize the calculation so as to minimize the chance of human error, rather than to derive formulas that look nice to the human eye.}, we have used
\be 
\label{w,y}
w\equiv e^{-\lambda/2}, ~~~y\equiv \lambda _2
\ee
 as the two unknowns. Then the algebraic relation can be written as
\be \label{al}
1-e^{-\nu /2}={F\over G}
\ee
where 
\beq 
F=(1+2c_2y+c_3y^2)(1-w^2+m^2r^2(2y+c_2y^2))  \nonumber \\
-4(1+c_2y)w(1-w),~~~
\eeq
\be 
G=4(c_2+c_3y)w(1-w)-m^2r^2(1+2c_2y+c_3y^2)^2.
\ee
and the final form of the two equations for the two unknowns follows from (\ref{e0n}, \ref{ebn}) and (\ref{al}): 
\be \label{f1}
a_1w'+b_1y'=d_1
\ee
\be
\label{f2}
a_2w'+b_2y'=d_2
\ee

Here (\ref{f1}) is just (\ref{e0n}) in the new notation, so that 
\be 
a_1=2rw
\ee
\be 
b_1=-m^2r^3(1+2c_2y+c_3y^2)w
\ee
\beq 
d_1=1-w^2+m^2r^2(2y+c_2y^2+ \nonumber \\
(1+2c_2y+c_3y^2)(1-w+wy))~~~
\eeq

To calculate the coefficients $a_2$, $b_2$, $d_2$, we first write eq.(\ref{ebn}) as 
\be 
\label{ebnn}
\alpha \nu'=\beta (1-e^{-\nu /2})+\gamma 
\ee
where 
\be 
\alpha =rw(1+2c_2y+c_3y^2)
\ee
\be 
\beta =4(1-w)(c_2+c_3y)
\ee
\be 
\gamma =4(1-w)(1+c_2y)
\ee

We then substitute $\nu$ and $\nu'$, as given by  \eqref{al}, into \eqref{ebnn}. We will need the derivatives:
\be 
\label{F'}
F'=F_1w'+F_2y'+F_3
\ee
\be 
G'=G_1w'+G_2y'+G_3
\ee
where 
\be 
F_1=-2w(1+2c_2y+c_3y^2)-4(1+c_2y)(1-2w) 
\ee
\beq 
F_2=2(c_2+c_3y)(1-w^2+m^2r^2(2y+c_2y^2))+ \nonumber \\
2(1+2c_2y+c_3y^2)m^2r^2(1+c_2y)-4c_2(w-w^2) \nonumber \\
\eeq
\be 
F_3=2m^2r(1+2c_2y+c_3y^2)(2y+c_2y^2)
\ee
\be 
G_1=4(c_2+c_3y)(1-2w)
\ee
\be 
G_2=4c_3(w-w^2)-4m^2r^2(1+2c_2y+c_3y^2)(c_2+c_3y)
\ee
\be 
G_3=-2m^2r(1+2c_2y+c_3y^2)^2
\ee

Then (\ref{ebnn}) and (\ref{al}) give 
\be 
a_2=2\alpha (GF_1-FG_1)
\ee
\be  
b_2=2\alpha (GF_2-FG_2)
\ee
\be 
d_2=-2\alpha (GF_3-FG_3)+(G-F)(\beta F+\gamma G)
\ee

For black holes, the $r$-derivatives in equations (\ref{f1},\ref{f2}) are converted to the $\rho$-derivatives by
\beq
w d/dr = d/d\rho\,.
\eeq

\renewcommand{\theequation}{C\arabic{equation}}
\setcounter{equation}{0} 

\section*{C. Static spherically symmetric field with matter}

To generalize the formalism of Appendix B in the presence of a perfect fluid one adds
\beq
{\cal T}^\mu_\nu=\rm{diag}(\ep,-p,-p,-p)\,,
\eeq
to the r.h.s. of Einstein equations which changes (\ref{e0},\ref{e1}) and (\ref{e0n},\ref{e1n}) accordingly. Besides, there are now two copies of stress-energy conservation \eqref{eb}, one is for the mass term \eqref{ebn} and the other one 
\be
\label{p'}
p'= -{1\over 2}\nu'(\ep+p)\,,
\ee
describing the hydrostatic equilibrium. The equation of state $p(\ep)$ and \eqref{p'} complete the system (\ref{e0n}, \ref{e1n}, \ref{ebn}), to be solved for five unknowns $\nu$, $\lambda$, $\lambda _2$, $\ep$ and $p$.

We still have two equations with $\nu'$ as the only derivative. This allows to express $\nu$ and $\nu'$ in terms of $\lambda,\lambda_2$, and $p$:
\be 
1-e^{-\nu /2}=\tilde F/ G
\ee
with
\be 
\tilde F= F + (1+2c_2y+c_3y^2)pr^2\,,
\ee
and 
\be
\label{nu'}
\nu'=\beta \tilde F/(\alpha G)+ \gamma/\alpha\,,
\ee
where $F$, $G$, $\alpha$, $\beta$ and $\gamma$ are given in Appendix B.

The rest of the calculation proceeds with minor modifications to yield
\beq
&\tilde a_1 w'+ \tilde b_1 y' = \tilde d_1\,,&\\
&\tilde a_2 w'+ \tilde b_2 y' + \tilde e_2 p' = \tilde d_2\,,&
\eeq
which can be integrated together with \eqref{p'}. Here the new coefficients are given by $\tilde a_1=a_1$, $\tilde b_1=b_1$, $\tilde a_2 = a_2$ and 
\beq
&\tilde d_1 = d_1-\ep r^2\,,&\\
&\tilde a_2 = 2\alpha (G\tilde F_1-\tilde F G_1)\,,&\\
&\tilde b_2 = 2\alpha (G\tilde F_2-\tilde F G_2)\,,&\\
&\tilde d_2 = -2\alpha (G\tilde F_3-\tilde F G_3)&\\ \nonumber
&~~~~~~+(G-\tilde F)(\beta \tilde F+\gamma G)\,,&\\ ~~~
&\tilde e_2 = 2\alpha r^2 (1+2c_2y+c_3y^2)G\,,&
\eeq
with
\beq
&\tilde F_1 = F_1\,,&\\
&\tilde F_2 = F_2 + 2(c_2+c_3y)pr^2\,,&\\
&\tilde F_3 = F_3 + 2(1+2c_2y+c_3y^2)pr\,.&
\eeq

\renewcommand{\theequation}{D\arabic{equation}}
\setcounter{equation}{0} 

\section*{D. The near-horizon solution}

We want to find analytic expressions for the near-horizon solution in the small-graviton-mass limit. We first find the  eigenvalues $\lambda_A$. Since $\lambda_1$ and $\lambda_2$ remain finite at the horizon, they can be replaced by their zero-graviton-mass limit. Substituting $e^\nu=e^{-\lambda}=1-r_g/r$ in the Bianchi identity \eqref{ebn} and solving for $\lambda_2(r)$ gives,  at $r=r_g$\footnote{The larger root was chosen for $\lambda_2$ because in the ``Vainshtein region'' $\lambda_2=1/\sqrt{c_3}$ which is larger than both roots.}:
\beq
\label{lambda2}
&\lambda_2=-2-c_2/c_3 +\sqrt{4+c_2^2/c_3^2-1/c_3}\,,&\\
&r_g\lambda_2'=-\lambda_2-2c_2/\sqrt{4c_3^2+c_2^2-c_3} \,,&
\label{lambda2'}
\eeq
here and below the prime denotes the $\rho$-derivative, with $dr/d\rho=w=e^{-\lambda/2}$. The same substitution $e^{-\lambda}=1-r_g/r$ in \eqref{l1} gives, at $r=r_g$,
\beq
\lambda_1=1+r_g\lambda_2'\,.
\eeq

Now we can find the desired solution. Due to finiteness of $\lambda_1$ and $\lambda_2$, the contribution of the mass term to \{00\} component of Einstein equations (\ref{e0n}) can be neglected. Thus we have
\beq
\label{w'}
2w'=(1-w^2)/r\simeq 1/r\,,
\eeq
which gives $w\simeq\rho/2r_{\min}$ and 
\beq
r \simeq r_{\min}+\rho^2/(4r_{\min})\,.
\eeq

Knowing $w$ and $y=\lambda_2$, the redshift $z=e^{-\nu/2}-1=-\lambda_0$ can be calculated from \eqref{al} and is of the form
\beq
z\simeq C_4/(\rho/r_{\min} +C_5 m^2r^2)\,,
\eeq
where $C_4$ and $C_5$ are positive dimensionless numbers for the parameter range of interest. 

Finally, to derive an expression for the curvature, we note that the near-horizon $T^\mu_\nu$ is dominated by the $\lambda_0\equiv -z$ contributions: 
\beq \label{lot1}
&T^1_1\simeq  m^2 z (1+2c_2\lambda_2+c_3\lambda_2^2)\,,&
\\
&T^2_2 \simeq  m^2 z (1+c_2(\lambda _1+\lambda _2)+c_3\lambda _1\lambda _2)\,.&
\eeq

Requiring that the leading order $T^1_1$ given by (\ref{lot1}) vanishes at the $\lambda _2$ value given by (\ref{lambda2}), gives 
\beq 
c_3=c_2^2
\eeq
And for $c_3=c_2^2$, 
\beq 
T^2_2 \simeq  m^2 z (1+c_2\lambda _1)(1+c_2\lambda _2)
\eeq
also vanishes at the $\lambda _2$ value given by (\ref{lambda2}).

\renewcommand{\theequation}{E\arabic{equation}}
\setcounter{equation}{0} 

\section*{E. The asymptotically flat metric}

To define the graviton's potential in Lorentz invariant theories of massive gravity one has to introduce a flat reference metric $\eta_{\mu\nu}$. In the unitary gauge ($x^\mu=\phi^\mu$) in which we work, the appearance of $\eta_{\mu\nu}$ breaks the reparametrization invariance of general relativity. Nevertheless the theory is still invariant under simultaneous transformations of the physical metric and the reference metric, and therefore, it is convenient (and always possible) to take the latter to be of the form:
\beq
\label{eta}
\eta_{\mu\nu}dx^\mu dx^\nu=dt^2-dr^2-r^2d\Omega^2\,,
\eeq
when dealing with a spherically symmetric setup. The most general static metric which respects the symmetries of the problem is 
\beq
\label{g}
ds^2=e^{\nu}dt^2-e^{\sigma}dr^2-2\xi dt dr - e^{\mu}r^2d\Omega^2\,,
\eeq
where $\nu, \sigma, \xi,$ and $\mu$ are arbitrary functions of $r$. 

Here we show that the requirement of asymptotic flatness, in the sense that at $r\to\infty$, $g_{\mu\nu}$ approaches the flat metric as defined in \eqref{eta}, forces $\xi$ to be identically zero.

The key observation, first made by Salam and Strathdee (1977), is that the identity $g_{tr}R_{tt}-g_{tt}R_{tr}=0$ which holds for the Ricci tensor of \eqref{g}, imposes a purely algebraic constraint on the components of the metric 
\beq
\label{constraint}
g_{tr}T_{tt}-g_{tt}T_{tr}=0\,,
\eeq
with $T_{\mu\nu}$ being the effective stress-energy tensor of the mass term. 

For the metric \eqref{g}, and arbitrary potential that is a function of the eigenvalues of the matrix $H^A_B$ in \eqref{H} one has $T_{tr}=\kappa g_{tr}$, with the coefficient $\kappa$ which is non-singular at $g_{tr}=0$. It follows that eq. \eqref{constraint} divides the solutions into two categories

(i) $g_{tr}=0$,

(ii) $T_{tt}=\kappa g_{tt}$.

Now consider large values of $r$, where the metric perturbations $h_{\mu\nu}\equiv g_{\mu\nu}-\eta_{\mu\nu}$ is small by the asymptotic flatness assumption. Then, for any non-linear completion of the Fierz-Pauli mass term $h_{\mu\nu}^2-h^2$, we have
\beq
T_{tr}=-\frac{1}{2}m^2h_{tr}(1+{\cal O}(h))\,,
\eeq
giving $\kappa =-m^2/2$ at large $r$. This excludes the branch (ii) as it leads to a finite value for $T_{tt}$. 

The branch (ii) is the Schwarzschild-de Sitter solution of Salam, Strathdee (1977) and Koyama et al (2011).

\end{document}